\magnification=\magstep1
\baselineskip=20truept
\overfullrule=0pt
\input epsf
\def\ia{\'{\i}}
\def\singlespace{\baselineskip=17.5pt \lineskip=0pt \lineskiplimit=-5pt}
\centerline{\bf Does a Good Conservation of Jacobi's Constant Imply
a }
\centerline {\bf Good Orbital Numerical  Integration?: I. Resonant Orbits. }
\vskip1cm
\centerline{J. Espresate$^{1}$, G. Castro Ben\ia tez$^{2}$}
\vskip3cm
\centerline {E-mail: julia@astroscu.unam.mx}

\vskip2cm

$^{1,2}$Instituto de Astronom\ia a, Circuito Exterior,

Ciudad Universitaria C.P. 04510 M\'exico D.F., MEXICO
\vskip3cm
\centerline{Submitted to {\it Icarus}}

\centerline{Submitted 17 January 2003}

\centerline{Revised 2003}

\centerline {Manuscript Pages: 27}

\centerline {Tables: 2}

\centerline {Figures: 11}
\vfill\eject
\noindent
{\bf Proposed Running Head}: { Numerical Performance and the Jacobi's
Constant Conservation.}

\vskip2.5cm
\noindent{\bf Editorial correspondence to:}

\noindent
Dr. Julia Espresate

\noindent
Instituto de Astronom\ia a, Circuito Exterior,

\noindent
Ciudad Universitaria 04510 M\'exico D.F., MEXICO

\noindent
phone (5255) 5622 40 14

\noindent
Fax:  (5255) 5616 06 53

\noindent
E-mail: julia@astroscu.unam.mx

\vfill\eject
\noindent {\bf ABSTRACT}
\vskip0.5cm
In this paper we show two examples of  numerical orbital
integrations (Planar Circular Restricted Three Body Problem) 
in which even though the conservation of Jacobi's constant
is near to 1 part in $10^8$, the integration proves to be wrong. That is,
for some particular cases the Jacobi's constant value is {\sl insensitive}
to very important qualitative (not quantitative) 
changes in the integrated orbit which are produced by numerical errors.
We provide specific recommendations to test and ensure 
that a numerical code 
is working properly, regardless the numerical method employed.
\vfill\eject

\noindent
{\bf I. INTRODUCTION}

The planar circular restricted three-body problem  (hereafter PCRTBP)
consists
of two bodies; a primary or central body (of mass $m_1$), a secondary
or perturbing body (of mass $m_2$), both moving in circular
orbits about their common center of mass, and a test particle moving
under the gravitational effect of the other two masses without 
affecting their motion. In this scenario neither the total orbital energy
nor the total angular momentum of the whole system 
are conserved since the test particle 
does not affect the motion of the other
two bodies.
In particular, 
the energy and angular momentum
of the test particle's orbit are not constants of the motion.
However, the dynamical
system still has an integral of the motion, the Jacobi's constant, $C$
which, in a coordinate system in which $m_1$ and $m_2$ are always at rest,
exclusively depends 
on the particle's instantaneous
position and velocity coordinates (see Eq. 2.5).

The planar circular restricted three-body problem
is an excellent
approach to a very wide variety of dynamical problems 
within the Solar System.
Some examples are the orbits of comets when only the gravitational
effects of the Sun
and Jupiter are taken into account; non-collisional and collisional
planetary rings affected by a planetary satellite; 
asteroid's orbits subject to Jupiter's gravitational perturbations, etc.
Since there is no analytical solution to this problem, it is 
most common to solve the problem numerically. 
That is, to integrate (using some numerical method)
the full Newtonian equations of motion
for the system (see Eqs. 2.2 and 2.4) 
given the initial conditions for the three bodies
and the secondary to primary
mass ratio.

When integrating an orbit numerically in the PCRTBP scenario, in many
instances there
is no way to find out if the resulting orbit is correct or not,
because we do not know the solution in advance.
The only way to determine if the code is performing
the integration correctly is by monitoring (at each
instant of time) the value of the Jacobi's constant which
(of course) should remain constant throughout the integration.
However, computers do not have infinite precision
and therefore as the integration proceeds errors accumulate.
Hence one can only expect that Jacobi's constant is conserved
to a {\sl reasonably large } number of digits.

We want to show that an integration
that gives a {\sl reasonably large} number of unchanged digits
in the value of Jacobi's constant, does not necessarily 
imply a good performance of the code.

Since all numerical methods accumulate errors, these results
are relevant to any numerical orbital integration
in the PCRTBP scenario, regardless the numerical method used for
the integration.

There is a large variety of numerical methods to integrate
ordinary second order differential equations, a very popular one
due to its high accuracy at low CPU cost is the Bulirsch-Stoer
method. The code that performed all the orbital integrations shown
in this paper is based on the implementation of this method that appears
in the 2nd. edition of {\sl Numerical Recipes} (Press {\it et al.} 1992).

\vskip1cm
\noindent
{\bf II. THE FULL NEWTONIAN EQUATIONS OF MOTION OF THE SYSTEM}

In what follows, the origin of the coordinate system is at
the center of the primary point-mass, $m_1$. We use
a Cartesian coordinate system to locate the position of
the perturbing point-mass $m_2$ and the test particle. 
In this reference frame the primary is always at rest;
the equation of motion for
the perturber $m_2$ is:

$${\bf \ddot r}_2=G(m_1+m_2){{\bf r}_2 \over r_2^3} \ \ \ . \eqno (2.1)$$

\noindent Where {\bf r$_2$} is the vector
position of the secondary with respect to the primary and
{\bf r$_2$} has time-dependent coordinates $(x_2,y_2)$.
The double dot denotes the second time derivative in all
of the equations, and $G$ is the gravitational
constant. Writing the components of Eq. (2.1) one obtains:

$$\ddot x_2 =- G(m_1+m_2){x_2 \over (x_2^2+y_2^2)^{3/2}} \eqno(2.2a)$$

$$\ddot y_2 =- G(m_1+m_2){y_2 \over (x_2^2+y_2^2)^{3/2}} \ \ \ .\eqno(2.2b)$$

\noindent
The equation of motion
for the particle is:

$${\bf \ddot r}=-Gm_1{{\bf r}\over r^3}
+Gm_2
\Bigg({{\bf r}_2-{\bf r}\over | {\bf r}_2-{\bf r}|^3}
-{{\bf r}_2 \over r_2^3} \Bigg) \ \ ,\eqno (2.3)$$

\noindent
where ${\bf r}$ with time dependent coordinates $(x,y)$,
is the position vector of the test particle with respect to
the primary.
The last term on the right
hand side of Eq. (2.3) is the {\sl indirect term}, which accounts
for the fact that the center of the primary
is not an inertial frame (e.g., Shu 1984; Murray and Dermott
1999) because it suffers accelerations due to its gravitational
interaction with the secondary.
Writing Eq. 2.3 by components 
one obtains:

$$\ddot x =- Gm_1{x \over (x^2+y^2)^{3/2}}
+Gm_2\Bigg\lbrack{x_2-x \over ((x-x_2)^2+(y-y_2)^2)^{3/2}}$$
$$-{x_2 \over (x_2^2+y_2^2)^{3/2}}\Bigg\rbrack \eqno (2.4a)$$

$$\ddot y =- Gm_1{y \over (x^2+y^2)^{3/2}}
+Gm_2\Bigg\lbrack{y_2-y \over ((x-x_2)^2+(y-y_2)^2)^{3/2}}$$
$$-{y_2 \over (x_2^2+y_2^2)^{3/2}}\Bigg\rbrack \eqno (2.4b)$$

Equations (2.2) and (2.4) form a coupled system of second
order differential equations. 
For this dynamical system the only constant of 
the motion is Jacobi's constant, $C$.
This constant is most
commonly defined in a frame that rotates with the secondary's mean
orbital frequency, $n_s$, and origin at the common center of
mass of $m_1$ and $m_2$. In  this frame it reads:

$$ C= n_s^2 ( \tilde x^2+ \tilde y^2) + 2\Big({\mu_1 \over r_1}
+{\mu_2 \over r_2}\Big)-\tilde v_x^2 -\tilde v_y^2 \ \ \ , \eqno (2.5)$$

\noindent
where $\tilde x$ and $\tilde y$ are the position coordinates
of the particle in this rotating frame, $\tilde v_x$ and
$\tilde v_y$ are the velocity components; $r_1$ and $r_2$ 
are the distances from the particle  to $m_1$ and $m_2$
respectively, and $\mu_1=Gm_1$ and $\mu_2=G m_2$.

Equations (2.2) and (2.4)  are input
into the numerical code in order to integrate in time, thus obtaining
the positions and velocities of both, the perturber and the test
particle at various time instants, $t$, such that $t=0 \leq t \leq t_f$
where $t_f$ denotes the time at which the integration is finished.
Note that our code integrates the equations of motion with the origin
at $m_1$ and includes the indirect terms. Hence the output
are positions and velocities with respect to an origin at
$m_1$ and the 
aforementioned Cartesian
axes. At this point it is worth to mention that the output coordinates
are then used to calculate the osculating elements for the particle's
orbit shown in all of the plots. Also, coordinates are transformed
into the rotating frame (origin at the center of mass) in order to calculate
Jacobi's Constant. All this transformations, may of course introduce
more errors, but this only makes our point stronger because we have
good conservation of Jacobi's constant and still the integration
proves to go eventually wrong.
\vskip1cm
\noindent
{\bf III. THE ORBITS}

To prove our point, we have chosen two different orbits
whose semi-major axes are near the 2:1 and the 3:2 Inner Lindblad
Resonances (ILR). 
In both cases the initial eccentricity of the test particle orbit is zero,
$e_0=0$,
and it starts at conjunction with the secondary, 
$\lambda_{p_0}=\lambda_{2_0}$, where the $\lambda$'s denote mean longitudes. 
For the two resonances consedered,
the initial semi-major axis of the test particle orbit is
very close to the {\sl nominal} location
of each resonance. In general, the nominal location 
(semi-major axis) of a first order resonance
is found through the commensurability of 
the mean orbital frequencies between the particle and the perturber
(or secondary). This means that the ratio of their orbital frequencies
(or periods) can be written as a ratio of two small integers,
that is:

$$n_p = {m\over m-1}n_2$$

\noindent where $m$ (integer)
indicates the number of the resonance;
$n_p=\sqrt{Gm_1/a_p^3}$ is the test particle's mean orbital
frequency and $n_2=\sqrt{G(m_1+m_2)/a_2^3}$ is the perturber's mean
orbital frequency. Solving for the semi-major axis of the
particle one obtains:

$$ a_n({m:m-1})= Gm_1\Bigg\lbrack {1 \over mn_2} 
\Bigg\rbrack^{2/3}\ \ , \eqno (2.5)$$

\noindent where subindex $n$ stands for {\sl nominal}, $a$ denotes
semi-major axis and the expression in parenthesis
on the left hand side is the usual
notation for a resonance. For the 2:1 ILR $m=2$ and for the 3:2 ILR $m=3$.

The  actual initial semi-major axes for the particle's orbit
chosen for the present work,
were found empirically
by Espresate and Lissauer (2001) when trying
to find the largest resonant forcing produced by the secondary
on the test particle. For the 2:1 ILR case,
this semi-major axis is given by:

$$a_p(2:1)=a_n(2:1)(1+ 1.5386878\times 10^{-4})\ \ \ ,\eqno (2.6) $$

\noindent  where $a_p$ was named, {\sl the perturbed
center of the resonance} in Espresate and Lissauer (2001)
(note the misprint in Eq. 3 of Espresate and Lissauer
2001). This semi-major axis is slightly
larger than the nominal and large perturbations occur.
With this initial conditions the particle's semi-major axis oscillates
in such a way, that its {\sl average semi-major axis} is the nominal
resonant semi-major axis hence, on average, the commensurability
of the orbital frequencies of the secondary and the particle
is closer to 2. If the particle starts at the nominal
semi-major axis its average semi-major axis is such that the frequencies
commensurability is farther from (integer) 2 
and hence the perturbations are smaller (see Espresate and Lissauer 2001).

For the 3:2 ILR case the {\sl perturbed center of the resonance}
was found at (Espresate 1997):

$$a_p(3:2)=a_n(3:2)(1 + 2.0565255 \times 10^{-4}) \ \ \ .\eqno (2.7)$$

Table I shows 
the initial conditions for the particle's
orbit in both cases; the semi-major axes are given in units of the distance
between $m_1$ and $m_2$ and are given with a large number of digits
that should all be taken into account.

\noindent {\bf [Table I ]}

As mentioned above our code uses the Bulirsch-Stoer
method, which (like most numerical methods), 
requires an input parameter called
$eps$ that represents the overall error tolerance and determines
the accuracy of the integration.

Since the code is written in double precision, we start 
using a very small tolerance, $eps$, which is
close to the machine double precision limit. For each of the subsequent
integrations we increase the value of $eps$ thus requiring
progressively less accuracy from the code. Table II shows
the $eps$ parameter,
and the length of each of the integrations, $t_f$, in units
of the perturber's orbital period. For a given ILR all of the
integrations have exactly
the {\sl same initial conditions}
for the particle's orbit.

\noindent {\bf [Table II ]}

\vskip1cm
\noindent
{\bf IV. RESULTS}

Figure 1 shows the results of
our best integration for the initial conditions in Table I (2:1 ILR, Run 1).
That is,
using the minimum error tolerance $eps=5 \times 10^{-15}$. From
top to bottom the plots are:
the semi-major axis in units of the separation between
$m_1$ and $m_2$ as a function of time (in units of the perturber's
period); second panel is the eccentricity. Note that its
maximum value coincides with the value given
analytically by Franklin {\it et al.} (1984) for a 2:1 ILR and the mass ratio
used in this work; third panel is the resonant argument,
$\phi=2\lambda_2-\lambda_p-\omega$ where $\omega$ is the instantaneous
periapse angle of the particle. As expected because the orbit
is inside the resonance region, the resonant argument librates
(as long as $e\neq 0$). Finally
the lowest panel shows the fractional conservation of Jacobi's constant
which by the end of the integration has a fractional change
a little over $\sim 10^{-10}$. Hence the constant is conserved
up to 9 digits approximately. It is interesting
to note that there is a time
interval in which the semi-major axis, the eccentricity,
and the resonant argument stay almost constant as it happens
for the 2:1 {\it exact resonance} according to Winter and Murray (1997).
Nevertheless, the particle falls out of exact resonance and
moves back to the libration region and later repeats the cycle.

\noindent {\bf [FIG.1]}

Figure 2 shows the same quantities as Fig. 1 but
for Run 2, an integration using a tolerance 10 times larger
than in Run 1 ($eps=5 \times 10^{-14}$). Still the integration
looks good although Jacobi's conservation is almost 4 times worst
than in Run 1.

\noindent {\bf [FIG.2]}

Figure 3  shows the same plots as Fig. 2 but for Run 3. Again we have
increased the tolerance by a factor of 10 but now there is a
very notable change in the behavior of the integration. The eccentricity
never makes it back to zero as in the previous integrations. It
remains oscillating between the maximum and half the maximum
and seems to continue like that. Jacobi's conservation however
is of 7 digits or near to 1 part in $10^8$ which is still
acceptable in most published works. It can be seen that when
the qualitative change occurs, Jacobi's constant does not show
any signal (whatsoever) that something has started to go wrong.
Since the change in the orbit is not quantitatively large, this is
not a surprise, but certainly the integration is qualitatively wrong.
In particular, the resonant argument does not seem to be librating
as clearly as in the previous two runs specially after the first cycle.

\noindent {\bf [FIG.3]}

Figure 4 shows the results for $eps=5 \times 10^{-12}$
in which a slow increase in the lowest points of the oscillations  
in eccentricity starts to become visible and also no libration of the resonant
argument is visible. Note that the number of maxima in the eccentricity
oscillations is 6.
The conservation of Jacobi's constant could still
be considered acceptable up to 6 digits approximately.

\noindent {\bf [FIG.4]}

Figures 5 and 6 are the results from Runs 5 and 6 and it really becomes
a disaster. Specially (of course) for Run 6 in which the tolerance is simply
too large and the lower limit of the eccentricity oscillation is frankly 
increasing. Note also that in Fig. 5 the number of maxima for the eccentricity
is 9 for the same time interval while in Run 4,
there are only 4 maxima. The conservation of Jacobi's constant is of
only 5 digits approximately. For Run 6 the number of maxima
in the eccentricity oscillation is 14 and the amplitude 
keeps decreasing.
In Run 6 one has a really bad conservation of Jacobi's
constant of up to 4 digits only.

\noindent {\bf [FIGS. 5 and 6]}

Figures 7, 8 and 9 from Run 7, 8 and 9
respectively, show exactly the same kind of deterioration
than that at the 2:1 ILR as one increases 
the $eps$ value. The same qualitative changes which again
are not reflected in the conservation
of Jacobi's constant. Also the  overall conservation seems to be
a little better for the same values of $eps$ than in the runs
for the 2:1 ILR case.

\noindent {\bf [FIGS. 7, 8 and 9]}

The question that arises now is if one can trust 
'blindly' an integration with $eps=5 \times 10^{-15}$. The answer of course
is {\bf NO}. Even with this small tolerance, errors keep
accumulating (however slowly) and eventually the code
will start doing things wrong. That is the purpose of Runs 10 and
11  in Table II.
The results of Run 10 are shown in Fig. 10. It  can be seen 
that eventually 
the integration suffers a qualitative change of the same
kind, but much later in time.
When this happens (close to $t=375,000$ orbits), Jacobi's constant is
being conserved up to 8 digits (!) which is usually viewed as a good
numerical performance. The change is not significant,
quantitatively speaking and that is why
Jacobi's constant does not show any significant
variation. Nevertheless the results are wrong because that is
not the correct orbital behavior.

\noindent {\bf [FIGS. 10 and 11]}

Finally, Figure 11 shows a long integration for the 3:2 ILR
which shows a very similar effect after  $1.7 \times 10^5$
orbits, which is earlier than for the 2:1 resonance using
the same $eps$ value.

\vskip0.5cm
\noindent
{\bf CONCLUSIONS}

In this paper we have shown orbital numerical integrations
for two resonant orbits in the (PCRTBP scenario) in which, 
although Jacobi's constant conservation may be as good
as 1 part in 10$^{8}$ the results are wrong. A qualitative change
in the orbit eventually
shows up. When the qualitative change occurs,
the eccentricity and the semi-major axis keep
oscillating near the correct values producing no 
significant quantitative changes in the value
of Jacobi's constant. It is the accumulation of errors during
the integration what causes this qualitative change in the integrated
orbit which turns out to be incorrect. That is, in both cases
at the largest accuracy the eccentricity always makes it back to
its initial value, and so does the semi-major axis (Fig. 1 and 2).
As one requires less accuracy from the numerical code in order
to speed the performance, this qualitative change shows up
earlier in the integration; the less the accuracy, the sooner
it shows up.

In other words, based on the Jacobi's constant conservation
and requiring 7 or 8 digits to remain unchanged, do not
guaranty that the orbit is being integrated correctly.
Qualitative changes may happen and will not show on the
Jacobi's constant value. 

Regardless the numerical method used,
all computers accumulate errors and therefore the behavior
for these two orbits should be similar. May be with a different
method (or different computer) the qualitative change
will happen later in time for a given $eps$ value,
or may be earlier in time. It depends upon the combination of
the type of computer itself and the numerical method employed.
However, most surely it will eventually occur.

In view of this results, we warmly recommend to all
people that make orbital numerical integrations
in the PCRTBP scenario,
particularly to those who use
the implementation of the Bulirsch-Stoer method in 
Press {\it et al.} 1992, the following:
$i)$ first determine to total time, $t_f$ of
the integration you plan to perform; $ii)$ use
the $eps$ value you will use for your integration
and $iii)$ with those quantities,
perform an integration using the initial conditions
of any of the two orbits
presented here. If during your whole
integration interval the eccentricity behaves as shown
in Fig. 1, then you maybe more confident that your method
and your computer are most probably working well during
the time interval of your interest. If not, try to reduce
the $eps$ value or the time length of your integration.

We recommend these two orbits because first of all we
already have their initial conditions, and also because
they are good ``cheaters'', they can give you a good
conservation of Jacobi's constant but still be wrong.

Therefore, if your code integrates properly during $t_f$
any of these two orbits (using an $eps$ of your choice), 
you may be a bit more confident
that during your time interval your code is doing things correctly.
Specially if you are not working near resonances, where 
errors are consistently amplified.
Needless to say that if you have enough time, you can check
your results, specially qualitative behavior of the orbit
for different values of $eps$ and check how reliable is
the number of digits you are conserving in Jacobi's constant
during the whole integration length. In other words, try to
find out first if the orbit you want to integrate is not
one good ``cheater'' like these two.

\vskip1cm
\noindent
{\bf ACKNOWLEDGMENTS}

We want to thank the Instituto de Astronom\ia a at Ciudad
Universitaria in Mexico City for all the support
given to this research and to Dr. Jorge Cant\'o
for the computer facilities that allowed us to perform
the numerical calculations. 
\vskip1cm

\centerline{\bf REFERENCES}
\vskip1cm

Espresate, J. 1997. {\it Dynamics of Planetary
Rings. } PhD. Thesis, Universidad Nacional Aut\'onoma de M\'exico
\smallskip
Espresate, J., and J.~J. Lissauer 2001.
Resonant Satellite Torques on Low Optical Depth Particulate Disks. II.
Numerical Simulations .
{\it Icarus.}{152}{29 - 47}
\smallskip
Franklin, F.~A., M. Lecar, and W. Wiesel 1984.
Ring particle dynamics in resonances.
In {\it Planetary Rings} (R.~J. Greenberg, and A. Brahic, Eds.),
pp.~562 - 588, Univ.~Arizona Press, Tucson
\smallskip
Murray, C.~D. and Dermott, S.~F. 1999.
{\it Solar System Dynamics,} Cambridge Univ. Press, Cambridge
\smallskip
Press, W.H., Teukolsky, S.A., Vetterling, W.T., and Flannery, B.P. 1992.
{\it Numerical Recipes in FORTRAN,}(2nd. ed.; Cambridge Univ. Press,
Cambridge
\smallskip
Shu, F.H., 1984. Waves in planetary rings. 
In {\it Planetary Rings} (R.~J. Greenberg, and A. Brahic, Eds.),
pp.~513 - 561, Univ.~Arizona Press, Tucson
\smallskip
Winter, O.~C., and C.~D. Murray 1997.
Resonance and chaos. I. First-order interior resonances
{\it Astron. Astrophys.}{319}{290 - 304}
\smallskip
\vfill\eject

\centerline{\bf TABLES}
\vskip1.5cm

\vbox{\halign to 5.5in{\quad  #\hfil\quad & \quad #\hfil \quad &\quad #\hfil
\quad & \quad #\hfil \quad & \hfil #\hfil \quad & \hfil #\hfil \quad \cr
\noalign{{\hbox {{\noindent \centerline{\bf TABLE I}}}
{\vskip0.3cm \noindent \centerline{\bf Particle's initial conditions}
\vskip0.6em \hrule \vskip0.4em}}}
$a_p/a_2$\hskip6pc & $e_0$\hskip2pc & $\lambda_{p_0}$\hskip1pc & 
$\lambda_{2_0}$ & \hskip1pc$m_1/ m_2$\hskip2pc& ILR\cr
\noalign {\vskip 0.2em \hrule \vskip1.1em}
$0.63005724618926$ &0.0&0.0
&0.0&$ 1  \times 10^{-6}$ & 2:1\cr
$0.76329951620447$ &0.0& 0.0
&0.0& $ 1 \times 10^{-6}$& 3:2\cr
\noalign {\vskip1.1em \hrule \vskip0.5em}}}

\vskip3cm

\noindent\hskip2cm
\vbox{\halign to 3.3in{\quad  #\hfil\quad & \quad #\hfil \quad &\qquad #\hfil
\quad &\qquad #\hfil
\quad  \cr
\noalign{{\hbox {{\noindent \hskip7pc {\bf TABLE II}}}
{\vskip0.3cm \noindent \hskip3pc{\bf Numerical Integrations data}
\vskip0.6em \hrule \vskip0.4em}}}
RUN No. & $eps$ & $t_f$ & ILR\cr
\noalign {\vskip 0.2em \hrule \vskip1.1em}
1 &$5\times 10^{-15}$& 40,000&2:1 \cr
2 &$5\times 10^{-14}$&40,000&2:1 \cr
3 &$5\times 10^{-13}$&40,000&2:1 \cr
4 &$5\times 10^{-12}$&40,000&2:1 \cr
5 &$5\times 10^{-11}$&40,000&2:1 \cr
6 &$5\times 10^{-10}$&40,000&2:1 \cr
7 &$5\times 10^{-15}$& 40,000& 3:2 \cr
8 &$5\times 10^{-13}$&40,000& 3:2\cr
9 &$5\times 10^{-10}$& 40,000& 3:2 \cr
10 &$5\times 10^{-15}$ & $1\times 10^6$ &2:1 \cr
11 &$5\times 10^{-15}$&$ 4 \times 10^5$& 3:2 \cr
\noalign {\vskip1.1em \hrule \vskip0.5em}}}

\vfill\eject

\noindent {\bf FIGURE CAPTIONS}
\vskip1cm \noindent
Figure 1. Shown are the
results from Run 1 (Table II) and the initial conditions
in Table I for the 2:1 case. From
top to bottom the plots are:
the semi-major axis in units of the separation between
$m_1$ and $m_2$ as a function of time (in units of the perturber's
period). The eccentricity $e(t)$ as a function of time; 
the resonant argument (in degrees)
$\phi=2\lambda_2-\lambda_p-\omega$, as a function of time. The vertical axis
has limits $[-180,180]$.
The bottom panel shows the 
fractional conservation of Jacobi's constant
which by the end of the integration has a fractional change
a little over $\sim 10^{-10}$. Hence the constant is conserved
up to 9 digits approximately.
\vskip0.4cm \noindent
Figure 2. Same plots as in Fig. 1 but from results of
Run 2 (Table II). Note in the bottom panel that Jacobi's
constant conservation is a little worse when 
the value of $eps$ is increased.
\vskip0.4cm \noindent
Figure 3. Same plots as in Fig. 1 but from results
of Run 3 (Table II). Note there is a qualitative change in the 
orbit due to the accumulation of numerical errors
which is not reflected in any way on the panel at the bottom,
where the fractional change of Jacobi's constant is shown.
At the end of the integration the Jacobi's constant conserves
to approximately 7 digits. Note also that now the eccentricity
reaches its maximum 5 times whereas in the previous two figures
it happens only twice.
\vskip0.4cm \noindent
Figure 4. Results from Run 4 (Table II)
displayed in the same way as
in previous figures. The number of maxima reached by the
eccentricity is now almost 7 and it actually does not make it
to 0.02, the maximum now
(not noticeable in this resolution) 
is slightly lower than 0.02.
\vskip0.4cm \noindent
Figure 5. Results from Run 5 (Table II). Now the eccentricity shows
9 maxima and just like in the previous two figures, the resonant
argument (third panel from the top) is circulating except
during the first cycle. The conservation of Jacobi's constant
is of approximately 5 digits.
\vskip0.4cm \noindent
Figure 6. Note the appearance of 14 maxima in the eccentricity 
(second panel from the top) and the minima are increasing.
However the whole amplitude of the oscillation in semi-major
axis and eccentricity are smaller than in Figs. 4 and 5.
This integration has a very questionable conservation of
Jacobi's constant of approximately 4 digits. 
\vskip0.4cm \noindent
Figure 7. Same as in Fig. 1 but for 
the 3:2 resonance case, Run 7 in Table I.
Panels show the same quantities as in the previous figures.
Note that the semi-major axis and the eccentricity have
a very similar qualitative behavior as for the 2:1 case
(see Fig. 1). However the amplitude and the period of the periodic
oscillations are smaller. Jacobi's constant is conserved to almost
one more digit than in the integration for the 2:1 case, using
the same value of $eps$ (see Fig. 1).
\vskip0.4cm \noindent
Figure 8.
Results from Run 8 with an error tolerance a 100 times larger
than that in Fig. 7. As in the 2:1 case, the integration is
qualitatively different, the eccentricity (second panel from the top)
never makes it back to zero as in Fig. 7 but still Jacobi's
constant is conserved to approximately 8 digits (!) and shows
no sign when the integration starts to go wrong. Also
note the increase in the number of maxima reached by the eccentricity.
\vskip0.4cm \noindent
Figure 9. Results from Run 9 with an error tolerance
1000 times larger than in the previous case.
The number of maxima in the eccentricity has increased
noticeably and shows the same behavior as that observed in
Fig. 6 for the 2:1 case. Still the value of Jacobi's constant
is slightly better conserved than in Run 6.
\vskip0.4cm \noindent
Figure 10. A long term integration (million orbits) for
the 2:1 case for the smallest error tolerance.
It can be seen that
approximately at $t \sim 3.75 \times 10^5$, the code starts
to go wrong in the same way as it did for Run 3
at $t \sim 1\times 10^4$. At this point,
Jacobi's constant is being conserved up to 8 digits, which is more
accurate than what is reported in the literature some times.
\vskip0.4cm \noindent
Figure 11.
A long term integration ($4\times 10^5$ orbits),
for the 3:2 ILR case with the smallest error tolerance.
The code starts to misbehave at $t\sim 1.7 \times 10^5$ orbits,
which is almost a factor of 2 sooner than for the 2:1 case
(see Fig. 10). Even though Jacobi's constant is better conserved
than in Run 10. The graph at the bottom shows
rapid oscillations (which are not observed in the same graph in Fig. 10)
that may indicate a numerical instability of some sort.

\vfill\eject
\vbox{\centerline{
\hbox{
\epsfxsize 16truecm\epsffile{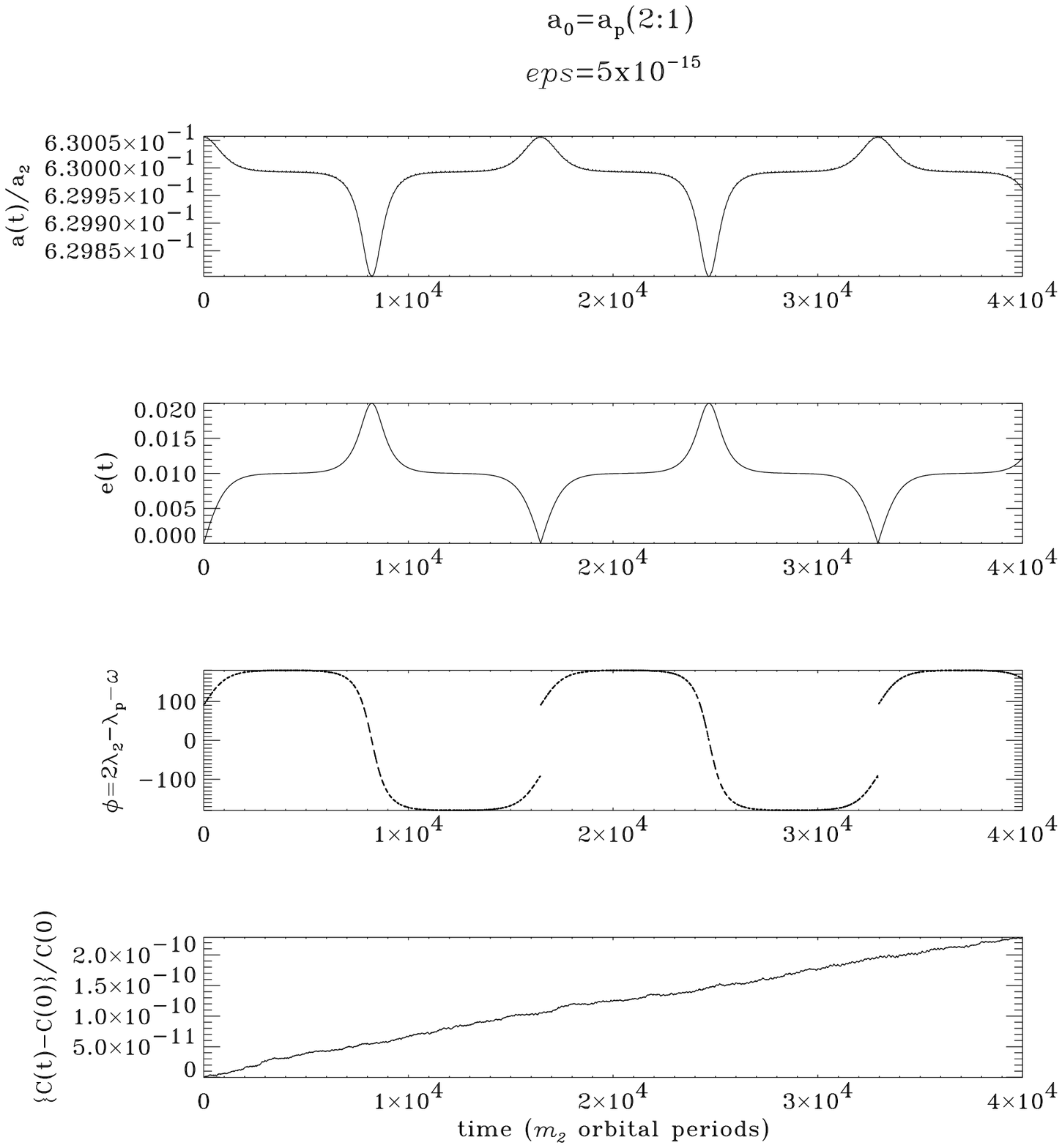}}}
\noindent \singlespace {{\bf Fig. 1}}}
\vfill\eject

\vbox{\centerline{
\hbox{
\epsfxsize 16truecm\epsffile{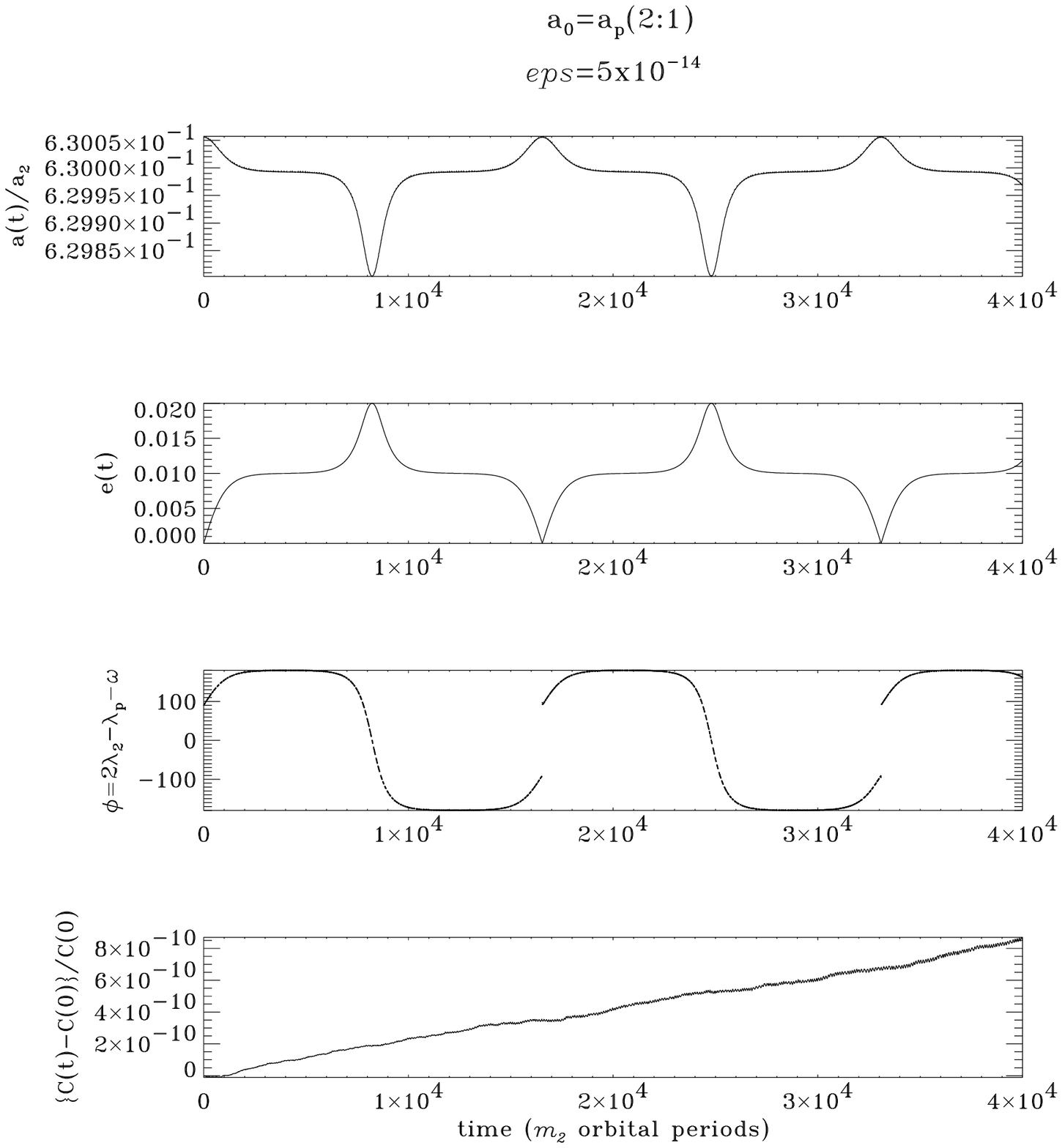}}}
\noindent \singlespace {{\bf Fig. 2}}}
\vfill\eject

\vbox{\centerline{
\hbox{
\epsfxsize 16truecm\epsffile{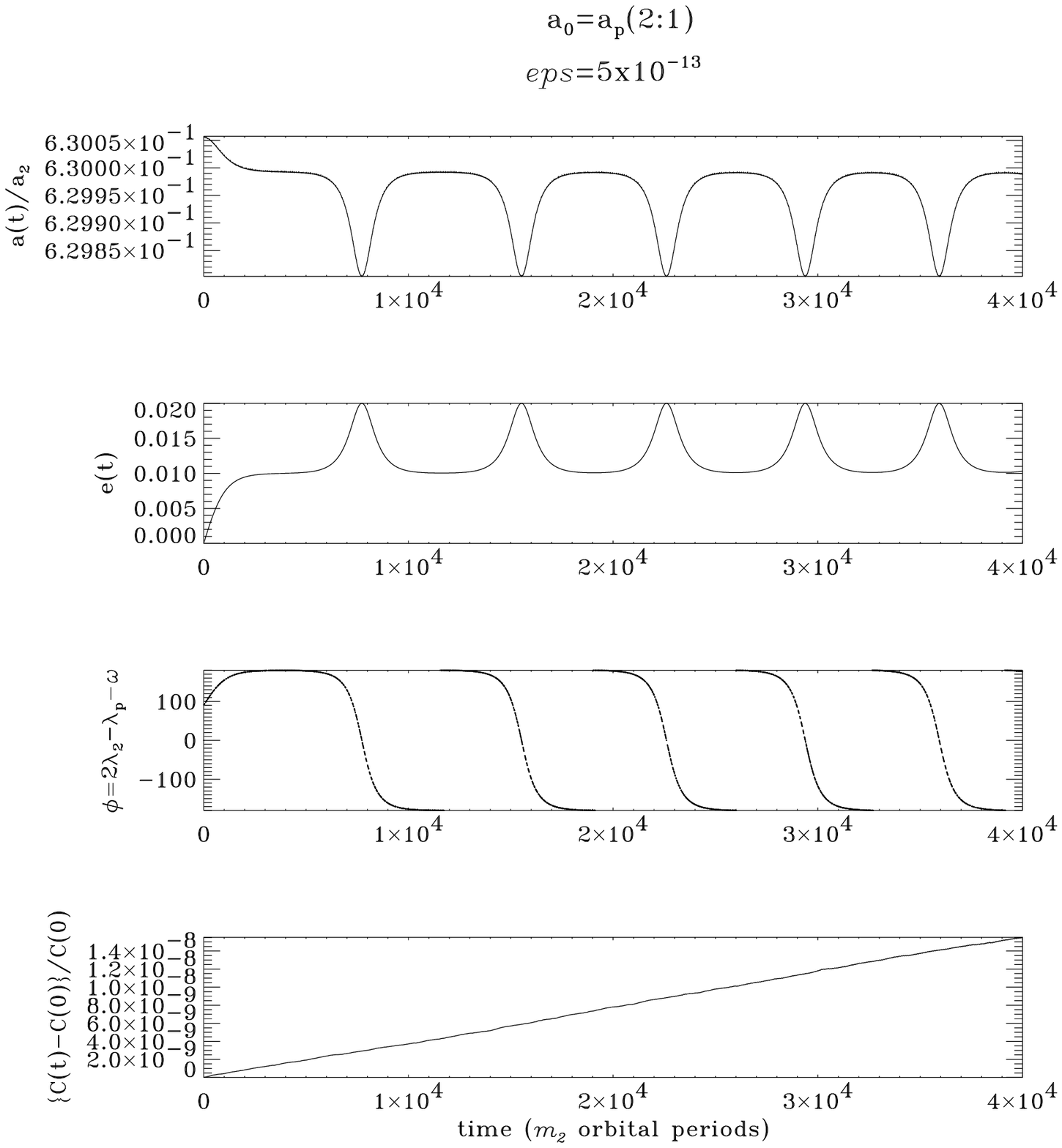}}}
\noindent \singlespace {{\bf Fig. 3}}}
\vfill\eject

\vbox{\centerline{
\hbox{
\epsfxsize 16truecm\epsffile{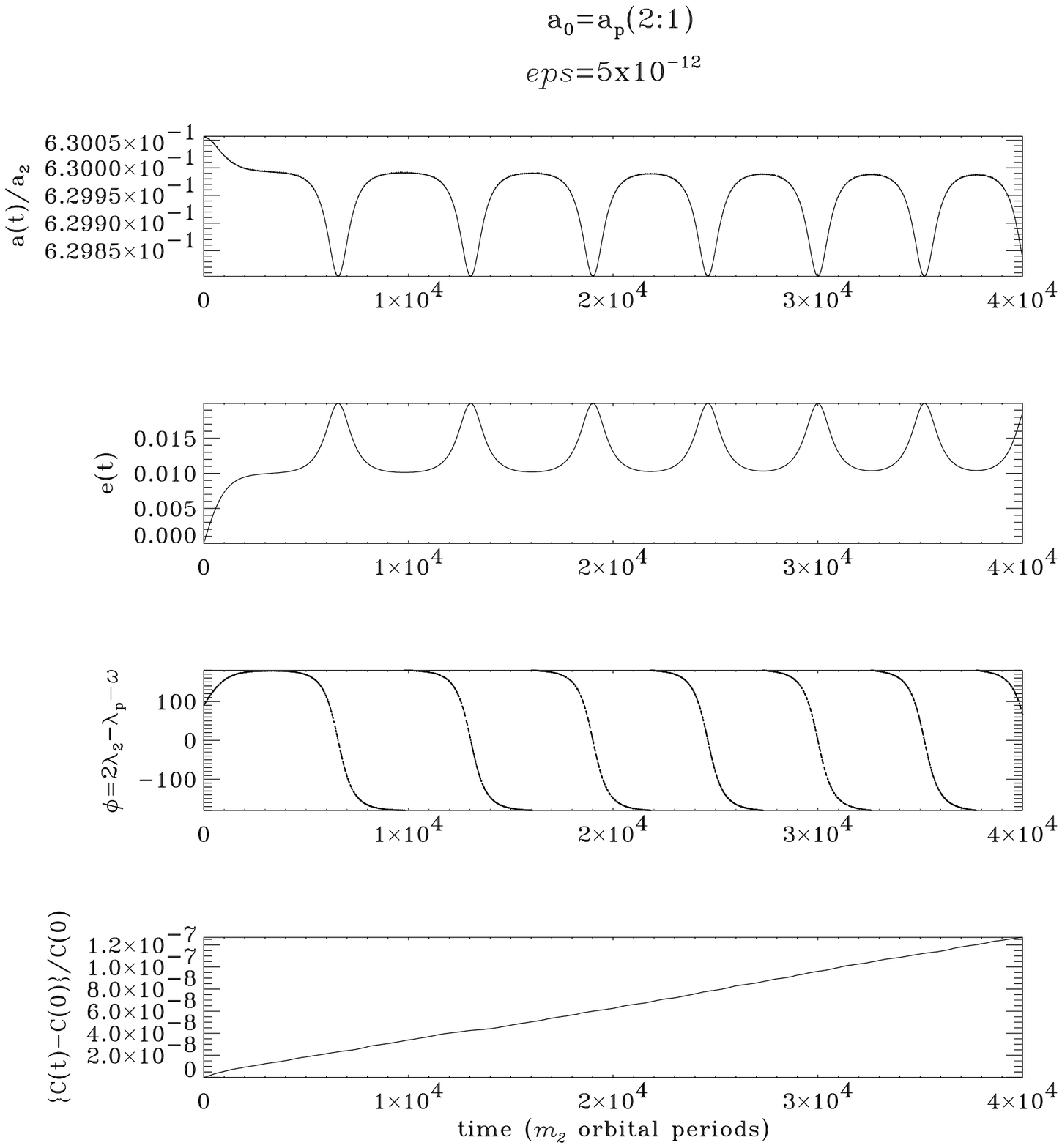}}}
\noindent \singlespace {{\bf Fig. 4}}}
\vfill\eject

\vbox{\centerline{
\hbox{
\epsfxsize 16truecm\epsffile{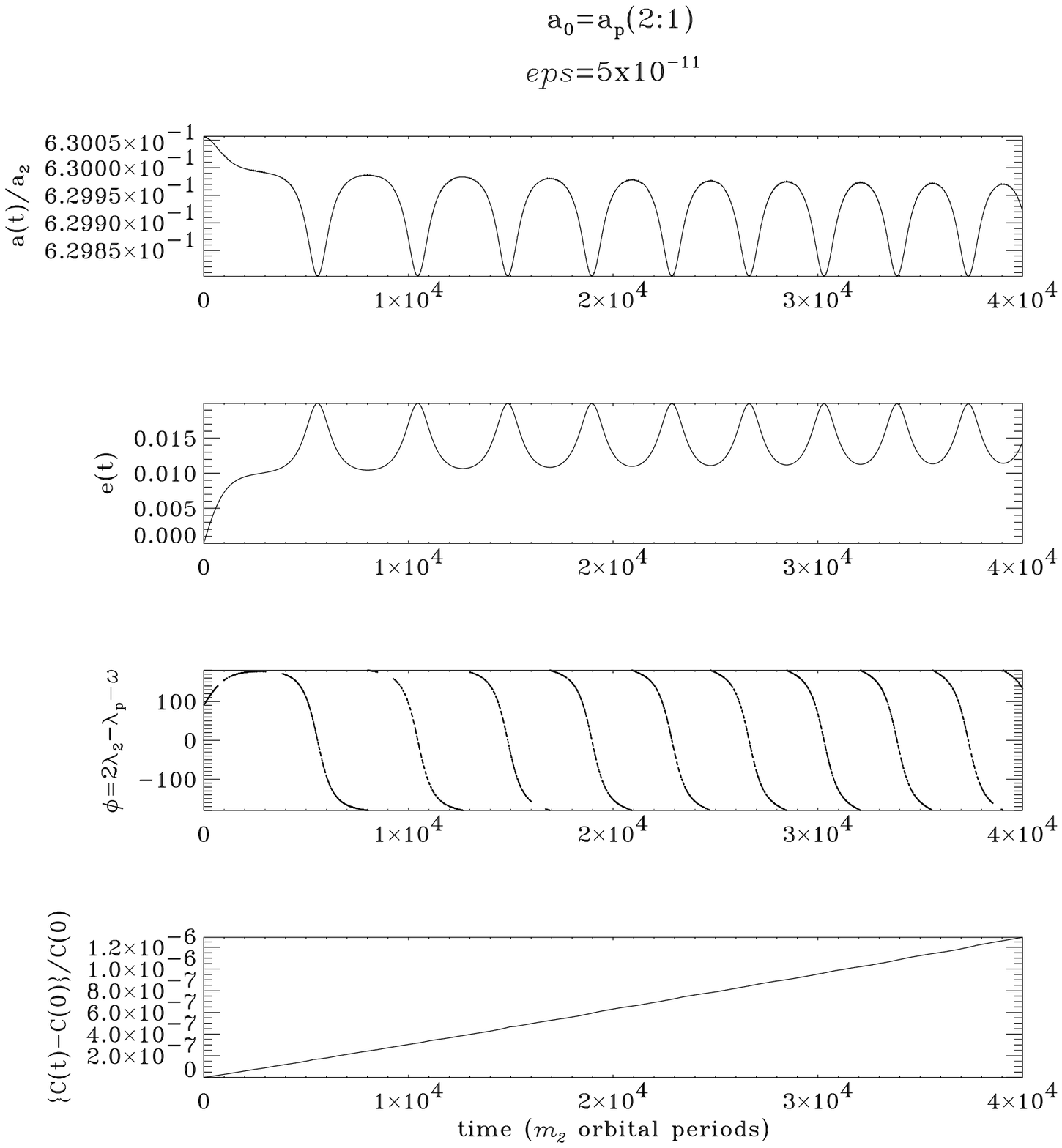}}}
\noindent \singlespace {{\bf Fig. 5}}}
\vfill\eject

\vbox{\centerline{
\hbox{
\epsfxsize 16truecm\epsffile{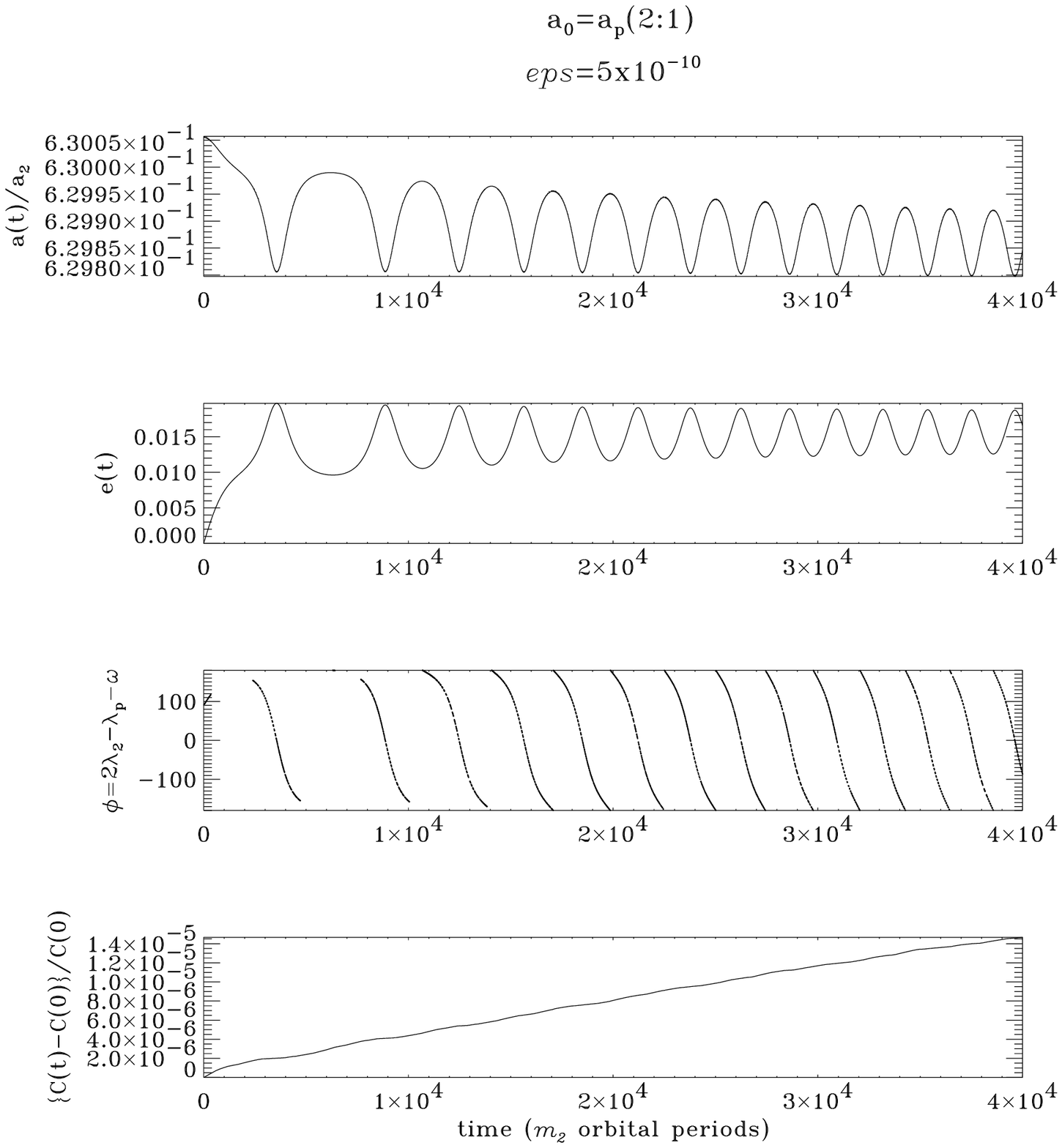}}}
\noindent \singlespace {{\bf Fig. 6}}}
\vfill\eject
\vbox{\centerline{
\hbox{
\epsfxsize 16truecm\epsffile{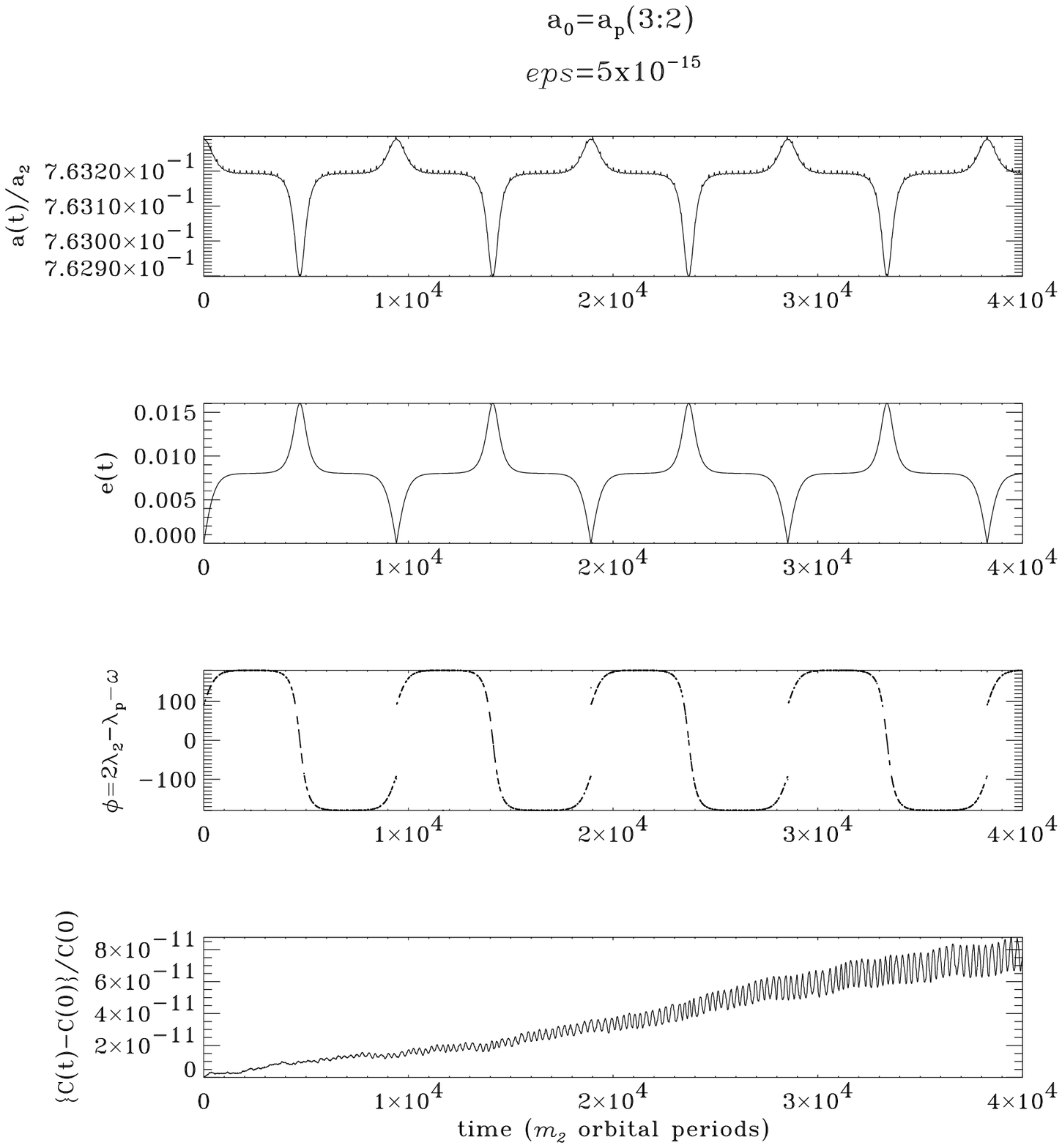}}}
\noindent \singlespace {{\bf Fig. 7}}}
\vfill\eject

\vbox{\centerline{
\hbox{
\epsfxsize 16truecm\epsffile{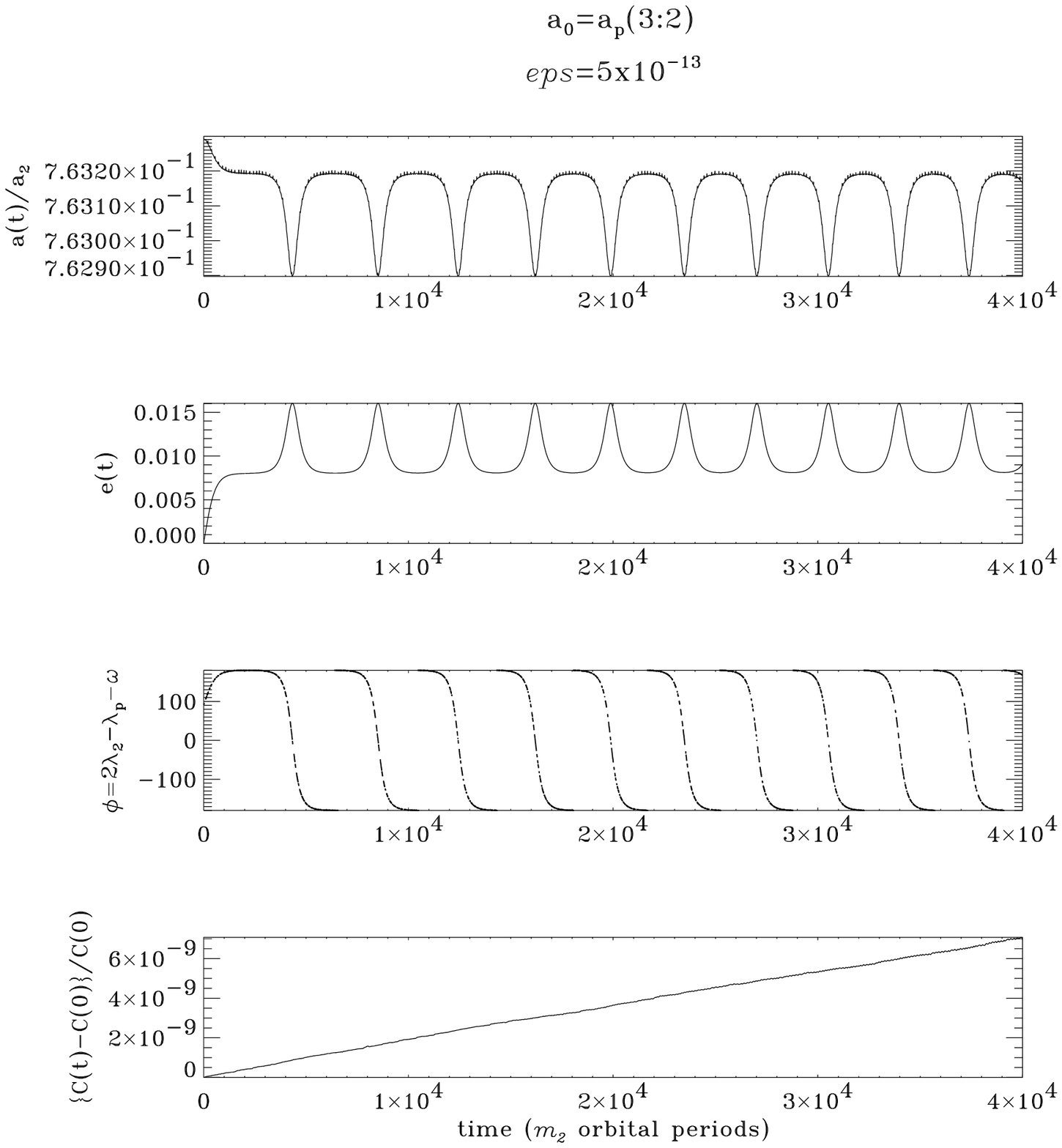}}}
\noindent \singlespace {{\bf Fig. 8}}}
\vfill\eject

\vbox{\centerline{
\hbox{
\epsfxsize 16truecm\epsffile{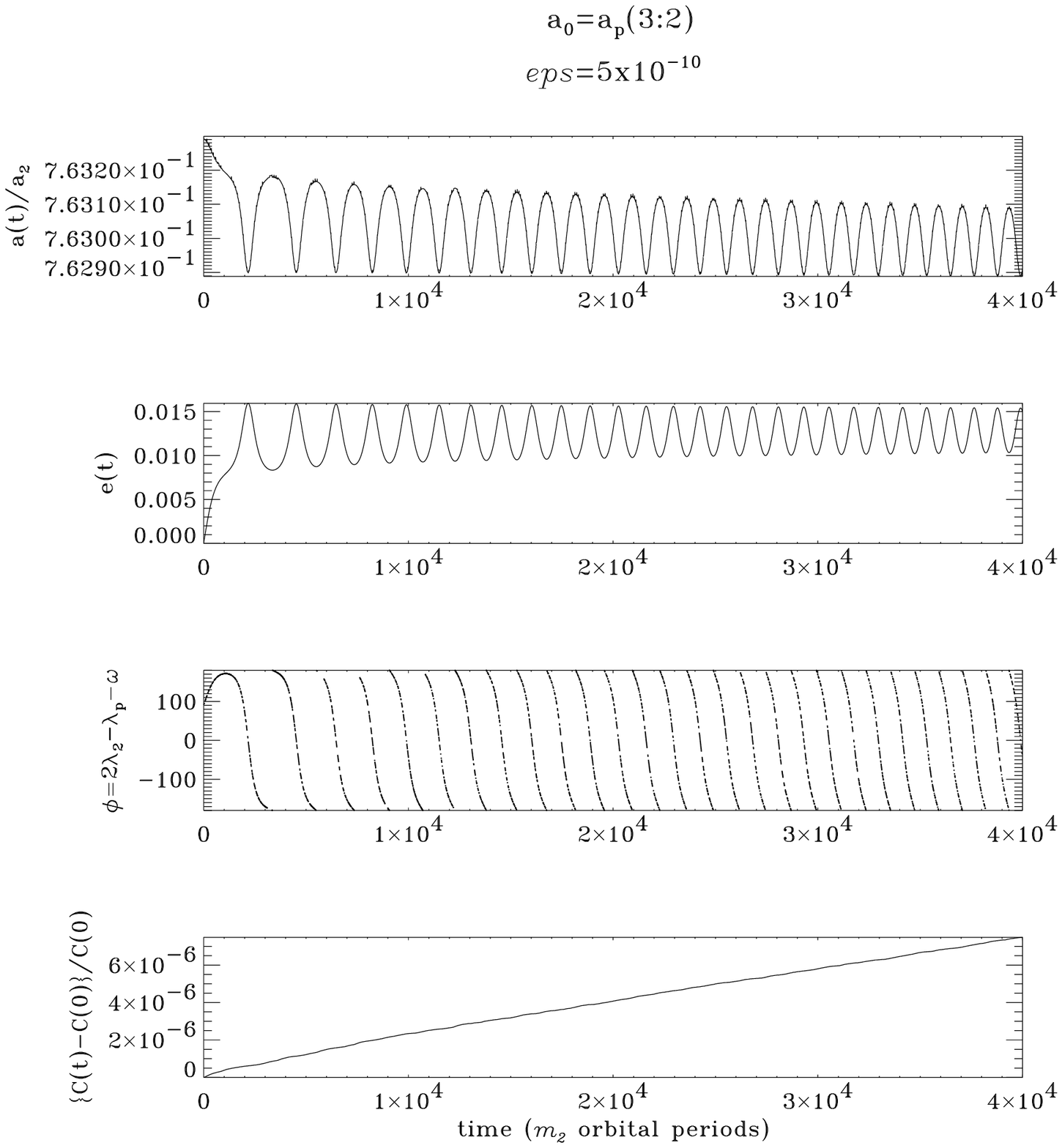}}}
\noindent \singlespace {{\bf Fig. 9}}}
\vfill\eject
\vbox{\centerline{
\hbox{
\epsfxsize 16truecm\epsffile{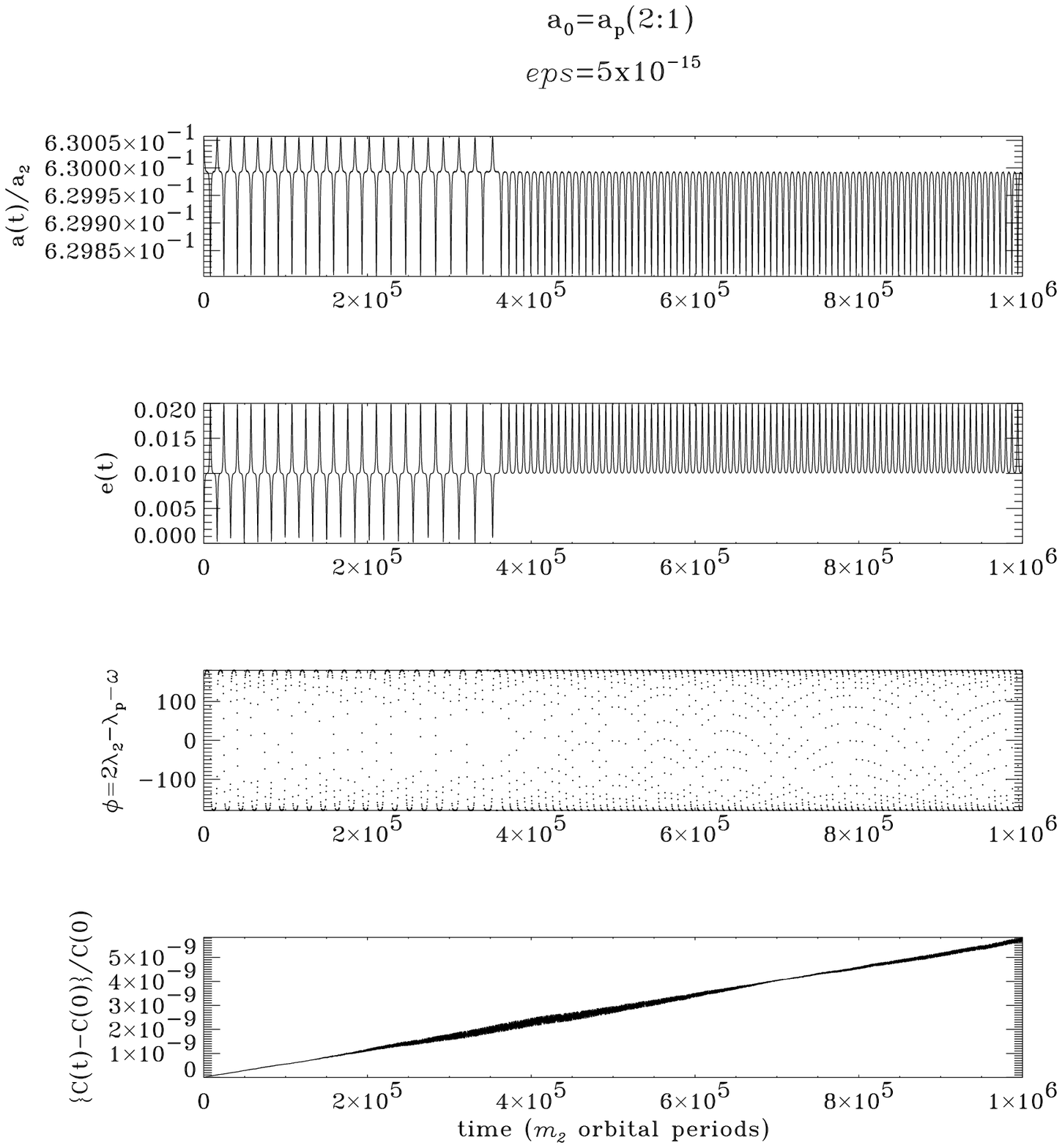}}}
\noindent \singlespace {{\bf Fig. 10}}}
\vfill\eject

\vbox{\centerline{
\hbox{
\epsfxsize 16truecm\epsffile{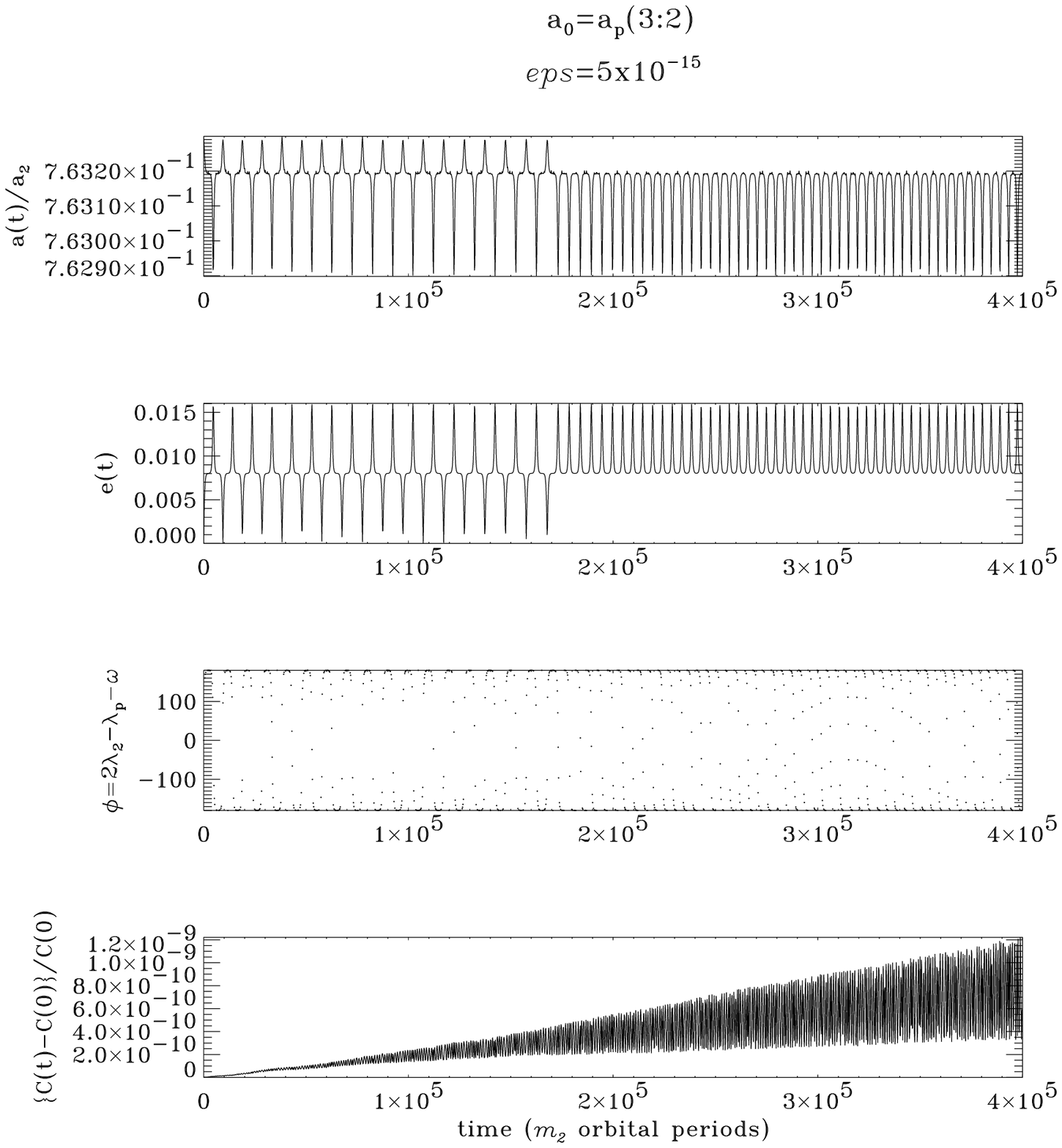}}}
\noindent \singlespace {{\bf Fig. 11}}}
\vfill\eject

\bye